Ostwald ripening controlled by diffusion of a sparingly soluble component

Alexey Kabalnov

Ink Splat Company, kabalnov@inkjet3D.com

## Abstract

Additives of sparingly soluble components are known to slow down or completely inhibit Ostwald ripening in dispersed systems. In this paper, our earlier model of stabilization against Ostwald ripening is revisited and extended. In a quasi-steady-state mode, the process is shown to be controlled by the diffusion of the less soluble component, and the whole machinery of the classical Lifshits-Slezov-Wagner (LSW) theory can be leveraged almost without any change. The particle size distribution is predicted to follow the same distribution function pattern as in the classic LSW theory. The rate of ripening follows the classic cubic law. To extend our earlier result, an improved extrapolatory equation for the ripening rate is derived, that covers the whole formulation range, accounts for the difference in molar volumes of the components and for the solution non-ideality.
The behavior described above is observed over the range of high concentrations of the poorly soluble component, with the cutoff determined by the lock-in number described in the previous paper of this series. When the concentration of the additive is low, the kinetics no longer follows the LSW pattern; instead, the particle size distribution becomes bimodal, with the fraction of 'fines' enriched by the poorly soluble component and the fraction of the large particles to ripen as if no additive were present. The lock-in parameter $L_1$ can be used to characterize for the transition from one mode to another. In the end, some practical stabilization approaches for emulsions are discussed.

## Introduction

Since the pioneering work of Higuchi and Misra [1], it was realized that an insoluble additive to the dispersed phase can slow down or eliminate Ostwald ripening in dispersions. In our earlier work, possible mechanisms of such stabilization were explored and the first qualitative theory developed [2]. Webster and Cates extended the model for the case of a completely insoluble additive [3].
In our previous paper of this series [4], we investigated the dissolution of a single drop onto a macrophase due to Laplace pressure effect. Three stages of dissolution were identified: the pre-lock-in stage, when the concentration of the poorly soluble additive undergoes as sharp increase, the lock-in stage, at which the Laplace pressure effect is nearly completely counterbalanced by Raoult effect, and the late lock-in stage, when the concentration of the poorly soluble additive undergoes a sharp increase, while still in the lock-in state. Whereas Ostwald ripening dynamics is considerably more complex than the case of dissolution of a single particle, some of the features of the processes are common, as we will be showing in this paper.
The structure of the paper is as follows. We start with the review of the LSW Ostwald ripening theory for one-component particles. We then explore in some detail the growth law for the particles in the two-component case. The case of zero solubility of the poorly soluble component is explored first. The possible scenarios are discussed as a function of the lock-in parameter $L_1$. For the high values of $L_1$, the system is predicted to enter the lock-in state, in which the particles come to equilibrium with each other and the mass transfer stops. On the other hand, when $L_1$ is small, the system is predicted to split to a bimodal distribution, with the fraction of fine particles being substantially enriched by the insoluble component, whereas the main fraction ripening as if no insoluble additive were present. The case of high values of $L_1$ is explored further and the condition of insolubility of the second component is relaxed to low, but finite solubility. For this, the mass flow of each component is analyzed and the coupling between them investigated, first on the level of chemical potentials, and then on the level of solubilities. Just as in case of dissolution of a single particle, the lock-in condition is predicted to emerge, with the rate controlled by the diffusion of the less soluble component. An extrapolatory equation for the rate, covering the whole range of compositions is derived. The lock-in number $L_1$, introduced in our previous paper for the dissolution of a single particle will be used in this case as well; however additional analysis is necessary because of the continuous range of the particle sizes, as $L_1$>>1 condition cannot be met for all the particles of the continuous particle size distribution. The paper is concluded by the summary and outlook for the stabilization of emulsions against Ostwald ripening.

**Ostwald ripening theory in one-component disperse phase case**

Ostwald ripening is the process of dissolution of small particles and the growth of larger once at their expense. With time, larger particles also differentiate in size; as a result, the average particle size in a system constantly increases . The thermodynamic driving force of the process is the reduction of the surface free energy. The underlying mechanism is based on the dependence of the solubility of the dispersed phase on the particle size, known as the Kelvin equation:

$$C(r) = \mathrm{C}_{\infty}^{*} exp\left(\frac{2\sigma V_m}{rRT}\right) = \mathrm{C}_{\infty}^{*} exp\left(\frac{\alpha}{r}\right) \approx \mathrm{C}_{\infty}^{*}\left(1+\frac{\alpha}{r}\right) \qquad (1)$$

Here *C(r)* is the solubility of the particle as a function of radius, *r*; $\mathrm{C}_{\infty}^{*}$ is the solubility of the macrophase, at the infinite radius of curvature, $\sigma$ is the interfacial tension, $V_m$ is the molar volume of the substance of the dispersed phase, *R* is the gas constant, and *T* is the absolute temperature. The Kelvin equation predicts a small increase in solubility compared to the macrophase. Thus, the parameter $\alpha = \frac{2\sigma V_m}{RT}$ is of the order of 1 nm for common emulsion systems; accordingly, it corresponds to 1% increase in solubility for 0.1 μm emulsion drops over the macrophase. The small magnitude of the effect allows one to expand the exponent in series the way it is done in eqn 1.

The Kelvin equation is based on the balance of the chemical potentials in the system. The excess Laplace pressure inside the particle generates and extra chemical potential of the solute, which produces the increased solubility:

$$\Delta\mu = \Delta p V_m = \frac{2\sigma V_m}{r} = RTln\left(\frac{C}{C_\infty^*}\right) \qquad (2)$$

Here the excess chemical potential, $\Delta\mu$, is measured with the respect to macrophase.

The theory of Ostwald ripening in the current form was developed by Lifshits and Slezov in 1958 [5] , and then, independently, by Wagner in 1961 [6]. It addresses the mass transfer in an ensemble of particles which is originally polydisperse and separated from each other by distances much larger than their sizes.
We start with the review of the original LSW theory. Consider growth/dissolution of a spherical particle in an infinite medium. The diffusion mass flux for a spherical particle in the steady state is equal to [7]:

$$J = 4\pi r D \Delta C \qquad (3)$$

$$\Delta \mathrm{C} = C(r) - C_M(t) \qquad (4)$$

Here $J$ is the mass flux, $D$ is the diffusion coefficient of the solute; the factor of $4\pi r$ can be considered as the ratio of the surface area of the particle $4\pi r^2$ to the effective diffusion path $r$ in spherical geometry. The subscript $M$ refers to the medium surrounding the particles and $C_M$ , which is the concentration in the medium at the infinity is time dependent, as, with the average size growth, the concentration in the medium also gradually decreases.
In a polydisperse system, larger particles grow at the expense of the smaller ones. At every moment of time there is a particle size that has a zero rate of change, which, in LS terms, is called the critical radius, $r_c$; at this radius, $C(r_c) = C_M(t)$. Equations 1 and 2 can now be re-written as:

$$J = 4\pi C_\infty^* D\alpha r\left(\frac{1}{r_c} - \frac{1}{r}\right) \qquad (5)$$

From this point on, the concentration units are reduced to the density of the solute and dimensionless; accordingly, the dimension of the flux is $cm^3/s$. For the spherical particle, growth law is as follows:

$$J = \frac{4\pi}{3}\frac{d}{dt}r^3 = 4\pi C_\infty^* D\alpha r\left(\frac{1}{r_c} - \frac{1}{r}\right) \qquad (6)$$

or

$$\frac{dr}{dt} = \frac{C_\infty^* D\alpha}{r^2}\left(\frac{r}{r_c} - 1\right) \qquad (7)$$

We now introduce the number particle size distribution function, *f(r, t)*. The growth rate law above can be incorporated into the continuity equation for the 'flow' of particles in the space of sizes:

$$\frac{\partial f}{\partial t} + \frac{\partial j}{\partial r} = 0 \qquad (8)$$

where $j = f \cdot \frac{dr}{dt}$ is the 'flux' of particles and $\frac{dr}{dt}$ is their 'velocity' in the space of sizes, per eqn (6).

Finally, the last equation is the mass conservation law:

$$\frac{d}{dt}\left[C_M(t) + \frac{4\pi}{3}\int_0^\infty dr\, r^3 f(r,t)\right] = 0 \quad (9)$$

Equations 6 - 8 is a complete set for the problem. In LSW theory, it is shown that with time it tends to the asymptotic solution of the following form:

$$f(r,t) \underset{t\to\infty}{\longrightarrow} n(t) \cdot\ P_\infty\left(\frac{r}{r_c(t)}\right) \cdot \frac{1}{r_c(t)} \qquad (10)$$

Here n(t) is the concentration of particles per unit volume and is equal to:

$$n(t) = \int_0^\infty f(r,t)dr \qquad (11)$$

LSW theory predicts the normalized distribution function *P(u)* where $u = r/r_c = r/\bar{r}$, as well as the time dependence of the critical radius and the particle concentration with time.
1. The cube of the critical radius (which happens to co-inside with the number-average radius $\bar{r}$ , at late asymptotic stage, see below) linearly increases with time per eqn 41. Accordingly, the number of particles per unit volume linearly decreases with time with the rate:

$$w = \frac{d(\bar{r}^3)}{dt} = \frac{8C_\infty^* \sigma V_m D}{9RT} \qquad (12)$$

2. The particle size distribution presented as a function of *u* with time becomes a time invariant:

$$P_\infty(u) = \frac{81 \cdot e u^2 exp\left[\frac{1}{2u/3-1}\right]}{\sqrt[3]{32}(u+3)^{7/3} \cdot (3/2-u)^{11/3}}, 0 \le u \le 1.5$$

$$P_\infty(u) = 0, u \ge 1.5 \qquad (13)$$

that is, it propagates forward in a self-similar way. The functional dependence is shown in Figure 1. This function is an 'attractor', that is, if initial distribution is wider, it becomes narrower with time, and vice versa, a narrower initial distribution expands to the attractor shape with time.

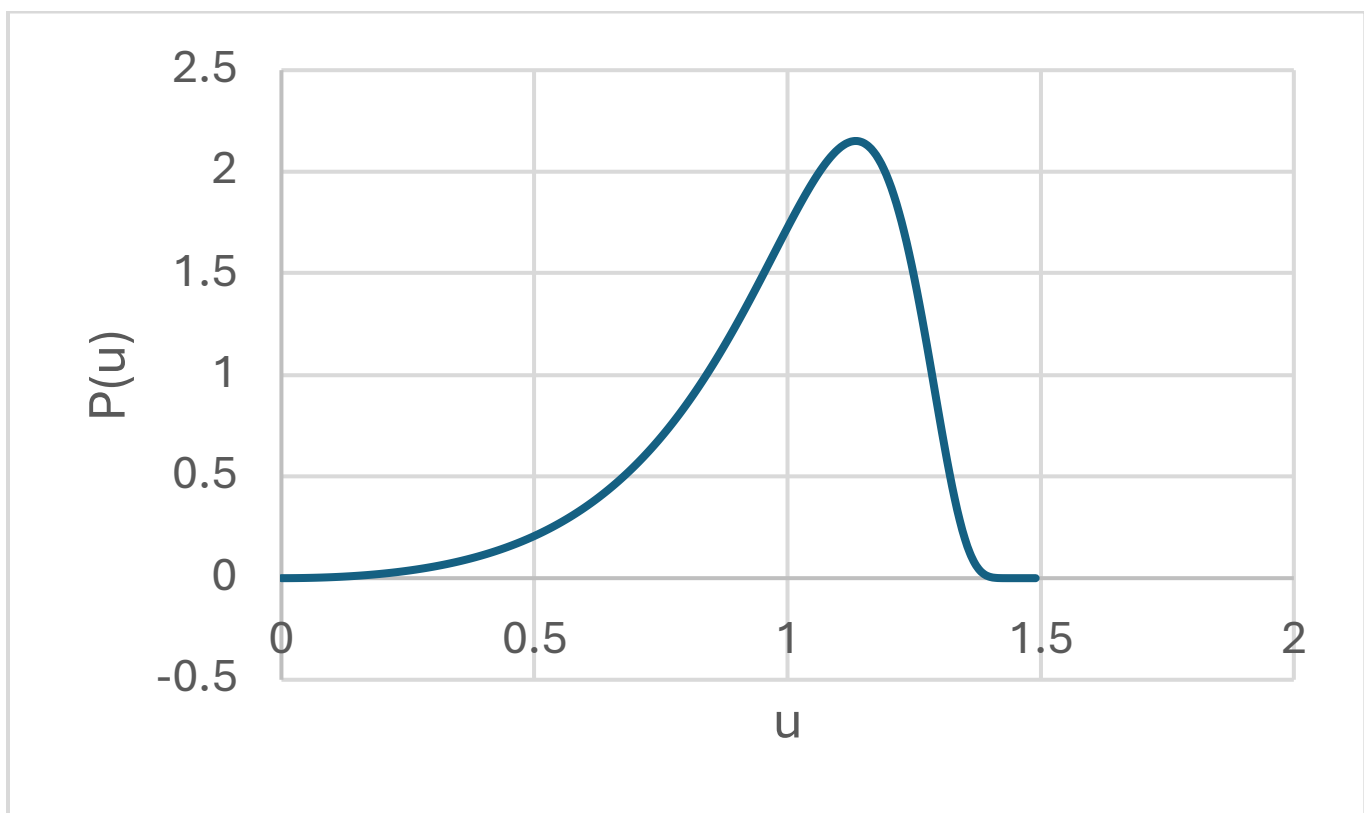

Figure 1. LSW particle size distribution attractor.

The asymptotic distribution function $P_\infty(u)$, as shown in Eqn (13) establishes at late times of the process . The transition time from the initial transient to the late stage asymptotic distribution depends on the initial particle size distribution in the system. In general, it takes longer for a wider distribution to shrink to the LSW attractor, than for a very narrow distribution to expand to it. Once the asymptotic regime is reached, the critical radius becomes identical to the number-averaged one, as the ratio of the first and the zeroth moment of the LSW attractor function is equal to 1. Before this late stage, however, the critical radius can be different from the number averaged value. Note that for the LSW attractor, the weight average radius is equal to $\bar{r}_w = \frac{M_4}{M_3}\bar{r} = 1.11\bar{r}$.

**Extension to two-component case**

We now will be exploring Ostwald ripening in a system with two-component dispersed phase [1]. Consider an ensemble of spherical liquid drops formed by two completely miscible components, 1 and 2. Both components are soluble in the medium, with the solubilities for the individual components $C^*_{\infty 1}$ and $C^*_{\infty 2}$, respectively. Note that those the values are for the pure components; the solubilities from the mixture are different, as the components reduce the solubility of each other; this analysis will be the subject of this study.

We assume that the second component is substantially less soluble in the medium by itself; $C^*_{\infty 2} << C^*_{\infty 1}$. Let the molar fractions of the components in the particles be $x_1$ and $x_2$, respectively. At time zero, the values are $x_{01}$ and $x_{02}$. We use the subscript 0 to indicate the time zero, that is, the composition of the drop before the mass transfer has started.

The steady state mass flows, $J_1$ and $J_2$ of the components are equal to[2]:

$$J_1 = 4\pi r D_1 \Delta C_1 \qquad (14)$$

$$J_2 = 4\pi r D_2 \Delta C_2 \qquad (15)$$

The transient effects to establish the steady state are neglected as shorter in duration. Also, coupling between the fluxes in this paper is approximated only on the level of chemical potentials of the diffusing components, that is, by how the chemical potential of one component affects another. Any other coupling is neglected. Finally, the concentration differences:

$$\Delta C_1 = C_1(r) - C_{M1}(t) \qquad (16)$$

$$\Delta C_2 = C_2(r) - C_{M2}(t) \qquad (17)$$

are the differences between the concentrations at the surface of the particle, and at the infinity in the medium, respectively. The concentrations at the particle surface $C_1$ and $C_2$ are size and time dependent, as both the particle composition and the Laplace pressure change with time. Partially, the dependence comes from the Kelvin equation, and partially because of the Raoult effect as the composition of the particle changes. The coupling between these fluxes is the main topic of this paper. We start with the case when the second component is completely insoluble in the medium and $J_2 = 0$, and all the mass transfer is due to the soluble first component.

### Insoluble second component: Scenarios of evolution for the particle size distribution

[1] Overall, the system has three components, the dispersion medium included, but the effects of the solubility of the medium in the dispersed phase is normally negligible and will be ignored.

To understand the evolution of the particle size distribution during ripening,  is instructive to explore the dependence of the chemical potential of the soluble component of a drop on the particle size, as a particle grows or dissolves. For ideal solutions, the excess chemical potential of the first component with respect to the state of the pure first component macrophase is equal to:

$$\Delta\mu_1 = RTln(1 - x_2) + \frac{2\sigma V_{m1}}{r} \qquad (18)$$

For a moment we consider the molar  of the second component small, $x_{02}$ <<1. This allows to expand the logarithm in series and the following equation can be obtained (Fig 2):

$$\frac{\Delta\mu_1}{RT} = \left(\frac{r_0}{r}\right)^3 x_{02} + \frac{\alpha_1}{r} \qquad (19)$$

Here $r_0$ is the initial radius of the particle just after it has been formed, $\alpha_1 = \frac{2\sigma V_{m1}}{RT}$. To remind, the parameter $\alpha_1$ has the dimension of length and for many emulsion and aerosol systems have the order of  1 nm.
We call the first term of the equation the Raoult term and the second term the Laplace term. As it can be seen from the graph, for larger sizes, the Laplace term is dominant, and the chemical potential  increases with decreasing the size in hyperbolic fashion, whereas for smaller sizes the Raoult term is dominant.  Although the graph describes the behavior of a single particle, it also provides the key to  evolution of a nearly monodispersed system with time.  It is of crucial importance to understand where on the graph the initial radius point  $r/r_0$ = 1 is located, that is,  is it before or after the maximum on the curve, and if the excess chemical potential positive or negative at this point. All these scenarios are possible, depending on the value of $x_{02}$, as discussed below.

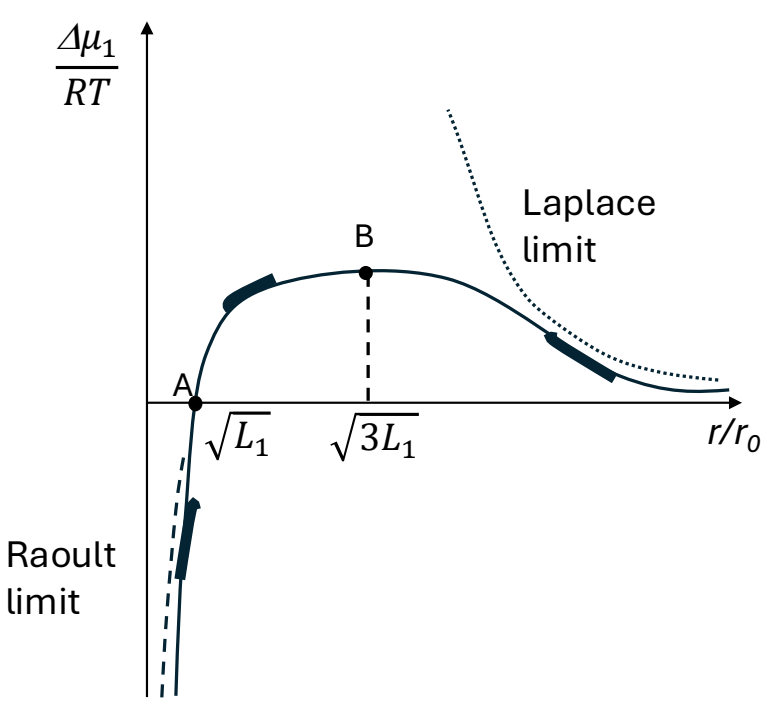

Figure 2. Cartoon illustrating the change in the excess chemical potential of the soluble component in the drop. Three possible locations for the initial radius are marked in bold, indicating, from left to right, the stable range (lock-in is possible), metastable range (lock-in is possible but susceptible to the system polydispersity) and unstable range (particles ripen).

If originally all the particles are located at the radii to the left of point A on the graph, the initial polydispersity of the particles does not cause the particles to diverge more in size, as there is a restoring force to bring them back to equilibrium; this is caused by the positive value of derivative of the chemical potential versus size. This range on the curve can be therefore interpreted as the range of the true lock-in, in the sense of our previous paper [4] . After the particles redistribute the soluble component between them, they come to complete equilibrium, which will hold if the second component is not soluble in the medium.

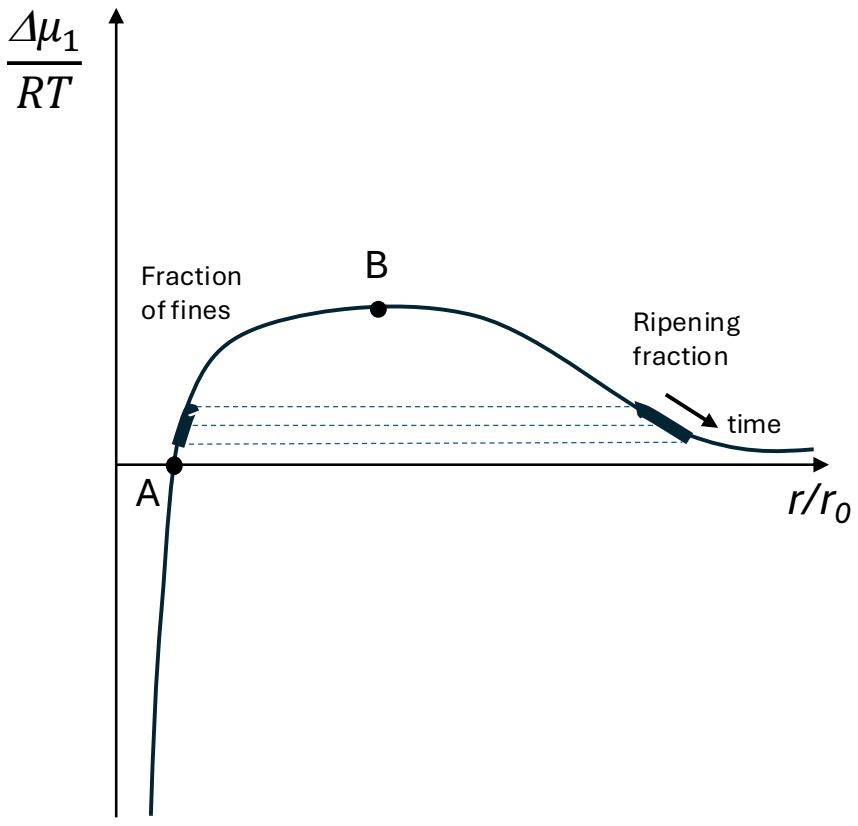

Figure 3. Cartoon illustrating the evolution of a ripening system with $L_1<1$. As the ripening proceeds in almost classical-style fashion, the ripening fraction gradually coarsens with time, and the excess chemical potential diminishes. All the small particle size drops are however caught into the fraction of 'fines', which co-exists with the ripening fraction. In the this process the total number of particles remains constant and does not decrease as $1/n \sim t$ as it would have been in case of the classical ripening.

On the other hand, if the particles are to the right of the point B, they cannot come to equilibrium with each other, as small differences in the particle size will only amplify with time due to the negative value of the derivative of the chemical potential versus size. As the particles are located on the Laplace branch of the curve, the system follows the classical Ostwald ripening scenario. As the system ripens, the average particle size increases, the supersaturation decreases and the excess chemical potential asymptotically tends to zero, and the particles gradually move downward on the curve (Fig 3). The main difference from the classical Ostwald ripening scenario is that the small particles cannot disappear completely; instead, they are caught in the AB branch of the curve. As the result, the system that was originally nearly monodisperse will be split into two fractions, the ripening fraction on the right, and the fraction of small particles on the left. The small particle size fraction will be technically in lock-in equilibrium with each other, whereas the large particle size fraction will not be in lock-in and will constantly ripen, gradually losing particles to the small particle size fraction.
Still another range worth considering is the range of initial sizes between the points A and B. If the particles have a narrow distribution that is all located between these points, they can enter the lock-in; however, as demonstrated by Webster and Cates [3], this equilibrium is metastable and if some polydispersity beyond this range will appear, the system will break into to fractions, as described above.

In our previous paper of this series, for the dissolution a single particle case, we introduced the dimensionless group called the first lock-in number , $L_1$:

$$L_1 = \frac{x_{02} r_0}{\alpha_1 (1 - x_{02})} \qquad (20)$$

where $r_0$ was the initial radius of the dissolving particle. Extending this approach to the case of a polydisperse system, going forward, we will replace $r_0$ with the number-averaged radius of the distribution, $\bar{r}$. We note that in Fig 2, Point A corresponds to the value of the lock-in parameter of $L_1=1$. In the following sections, we will conduct numerical simulations to validate the predictions of the model for the cases of $L_1 > 1$ and $L_1 < 1$ and various initial distribution functions, all done for the case of insoluble second component. We start with the hypothetical case of a uniform distribution, followed by the cases of log-normal and gamma-distributions, common in emulsion technology as the initial states just after dispersion [8], [9],[10]. We also analyze the evolution of the LSW attractor function, as this distribution can emerge during ripening; our interest will be by how much it will be affected by the transition into the lock-in.

**Insoluble Second Component, $L_1$>1, Lock-In State: Numerical Modeling**

At time zero, right after dispersion, all the particles have the same composition. The capillary pressure inside smaller particles makes the components to dissolve and diffuse into the larger ones. The more soluble Component 1 quickly diffuses out whereas Component 2 remains 'trapped'. This continues until the Raoult effect due to the increased concentration of the Component 2 in the smaller particles completely counterbalances the capillary driving force of the mass transfer. After the first component gets re-distributed, its chemical potential becomes equal for all the particles: the value settles at some $\Delta\mu_{1e}$ at which the mass transfer stops[3]:

$$\Delta\mu_1 = \frac{\partial\mu_1}{\partial x_2}\Delta x_2 + \frac{\alpha_1}{r}RT = \Delta\mu_{1e} = const \qquad (21)$$

This condition holds for all the particles in the initial distribution. Thus, there is a particle size in the middle of the initial range, which we call $r_e$, that neither grows nor diminishes; the exact location of $r_e$ in the distribution depends on the material balance in the system, as will be shortly illustrated below.

We conducted numerical simulations to illustrate how a dispersed system enters the lock-in state. In the first example, we explore the evolution of a uniform stepwise distribution of particles bracketed by $r_{min}$ = 100 nm and $r_{max}$ = 300 nm, with the number average radius $\bar{r}$ of 200 nm. The solution of the components in each other is assumed to be ideal with the value of $\alpha_1 = 10^{-7}$ cm, which corresponds to the values of parameters $\sigma$ = 10 dyn/cm, $\bar{V}_m$ = 80 $cm^3$/mol and T = 300K[4] . When the molar fraction of the insoluble component is high, $x_{02}$ = 0.1, the particle size distribution barely undergoes any change (Figure 4); for $x_{02}$ =0.01, however, the changes are significant. Figure 5 shows the relative change in the molar fraction of the insoluble component across the sizes; again, at $x_{02}$ =0.1 the change is minor, but it is significant at $x_{02}$ = 0.01.Note that for the cases of the uniform distribution discussed above, $\bar{r}$ = 200 nm and the values of $L_1$ are equal to 22 and 2, respectively.

The evolution of the stepwise distribution has been discussed merely for illustrative purposes; it would be of interest to check the evolution in more practically significant cases. Figures 6,7 shows the evolution of the LSW attractor function at the number average radius of 100 nm. Similar trends are observed; at $x_{02}$ =0.1 the change is minor, but it is significant at $x_{02}$ = 0.01. An important difference, however, is that the LSW distribution includes the range of ultimately small sizes, and, in the lock-in condition, the molar fraction of the insoluble component increases sharply in the part of the distribution where the radius tends to zero (Figure 7).

Similar calculations were also performed for log-normal, and gamma-function distributions that are common in emulsion manufacturing as the initial states [9], [10], [11]; very similar trends were observed, with minor changes to initial distributions at $x_{02}$ =0.1 , $L_1$ = 11, and much stronger changes at $x_{02}$ = 0.01, $L_1$ = 1. In both cases, even for $L_1$ = 11 case, a sharp increase in the concentration of the insoluble component is observed at very small particle sizes, see Figures 8-11.

---

[3] From this point on, we measure the excess chemical potential with respect to the non-dispersed phase, not the pure first component.

[4] More details of numerical simulation are given in the Appendix.

This sharp increase requires more discussion as the approximation that will be used below assumes the smallness of the relative changes in the particle composition over the particle lifetime:

$$\frac{\Delta\phi_2}{\phi_{02}} \sim \frac{\Delta x_2}{x_{02}} << 1 \qquad (22)$$

This condition can be met for distributions that are bracketed by $r_{min}$ and $r_{max}$ but cannot be met for continuous distributions such as the LSW attractor, or log-normal distributions. The argument that is made in this paper is that whereas this condition is indeed violated for small radii, it does not affect the particle lifetime significantly insofar as $L_1$ >>1. This is partially illustrated by Table 1, in which the volume fraction of this high concentration state of the second component is tabulated. As the lifetime of particles scales by the linearity of volume with time, the expectation is that the total effect on the particle lifetime will be minor, partially in line with the arguments of the previous paper [4] in which the effects of the 'late lock-in' on the particle lifetime was evaluated as minor for $L_1$>>1.

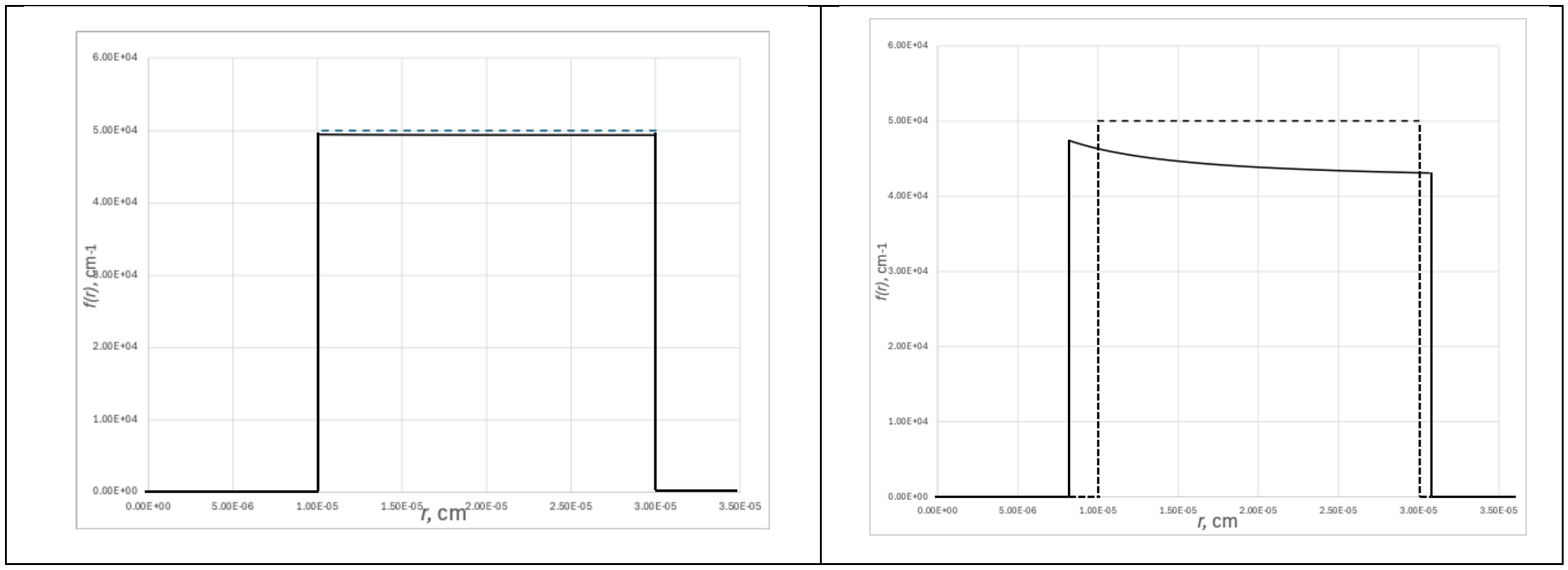

Figure 4. Evolution of a stepwise uniform number particle size distribution during the transition to lock in. Dashed lines: initial distributions; continuous lines: final distributions. Left: $x_{02}$ = 0.1;$L_1$ = 22; $\frac{r_e}{\bar{r}} = 1.25$ ; right: $x_{02}$ = 0.01, $L_1$ = 2, $\frac{r_e}{\bar{r}} = 1.20$.

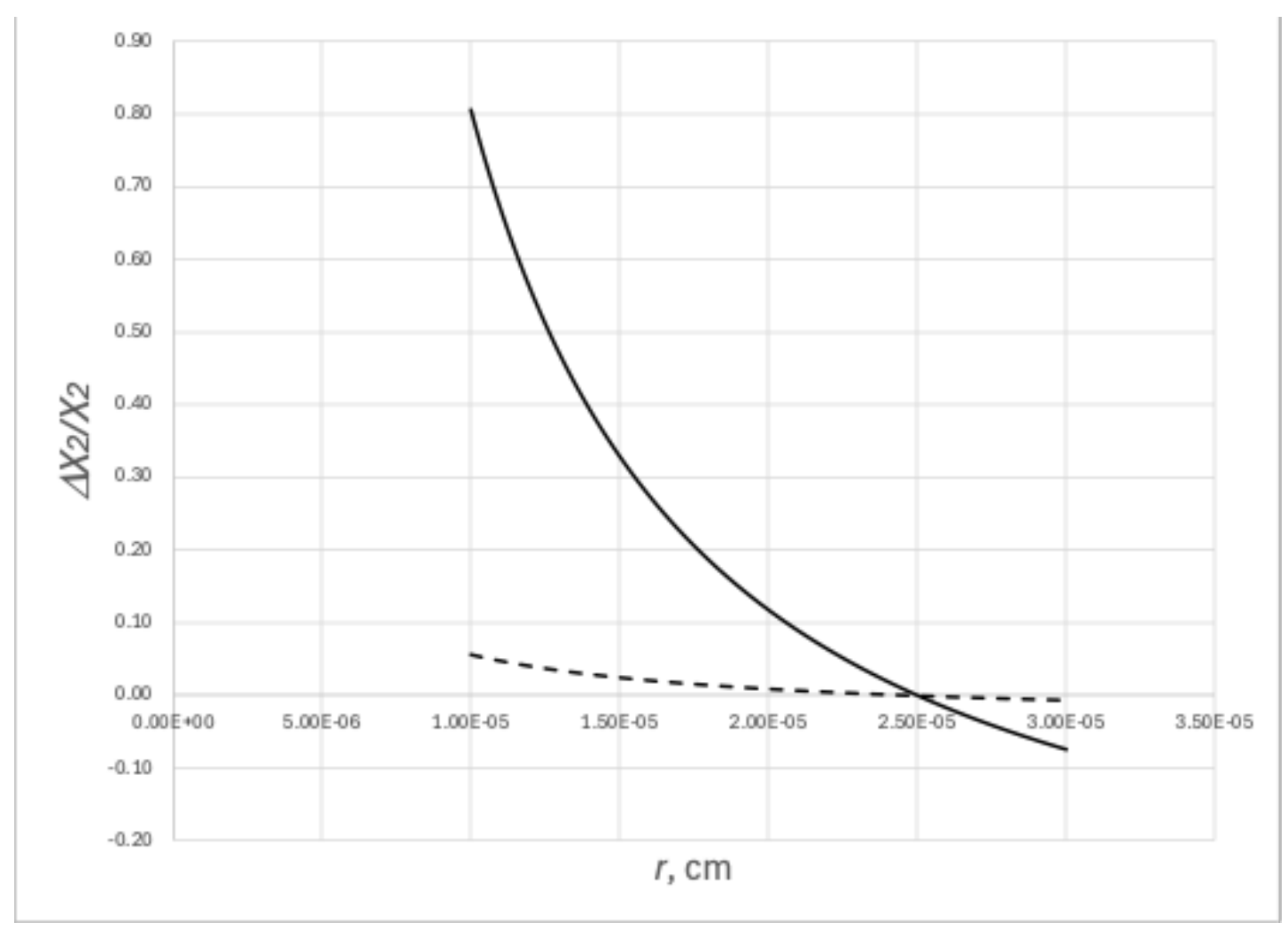

Figure 5. Relative changes in the initial value of the initial mole fraction $x_2$ during the transition to lock -n plotted versus the final particle size for the uniform distribution. The same set of parameters as in Figure 4; Dashed line: $x_{02}$ = 0.1, $L_1$ = 22; continuous line: $x_{02}$ = 0.1, $L_1$ = 2.

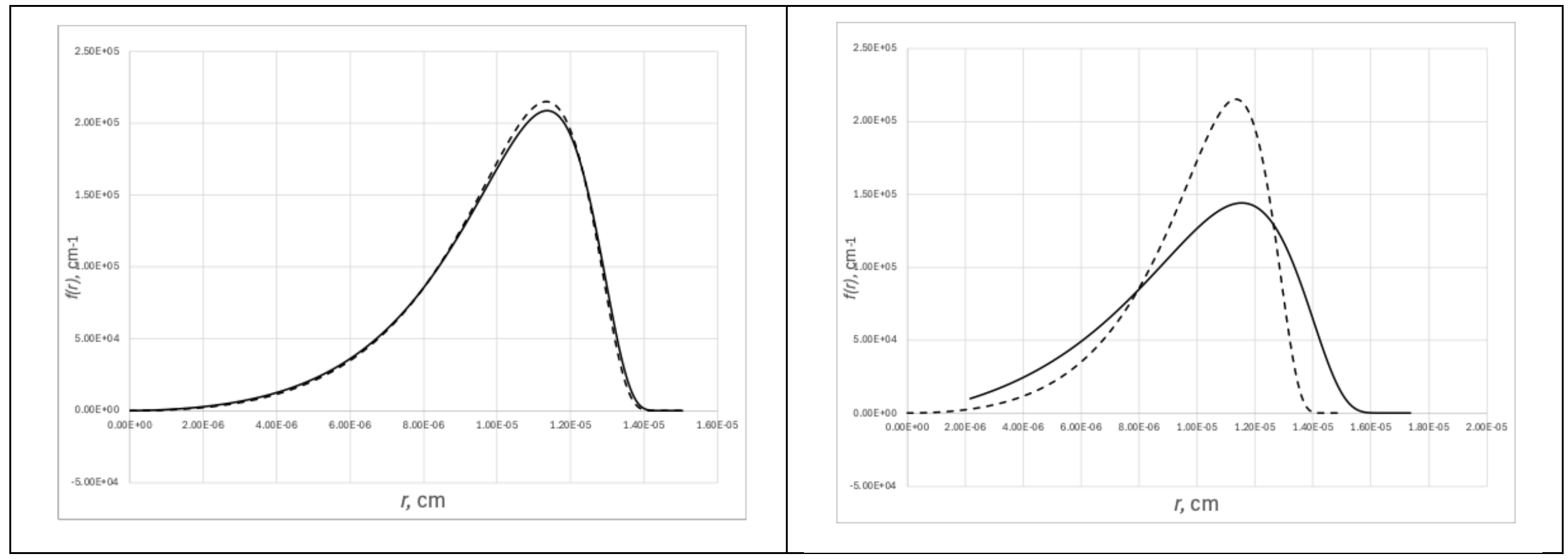

Figure 6. Evolution of LSW number particle size distribution during the transition to lock in. Dashed lines: initial distributions; continuous lines: final distributions. left: $x_{02}$ = 0.1;$L_1$ = 11; $\frac{r_e}{\bar{r}} = 1.14$; right: $x_{02}$ = 0.01, $L_1$ = 1, $\frac{r_e}{\bar{r}} = 1.21$.

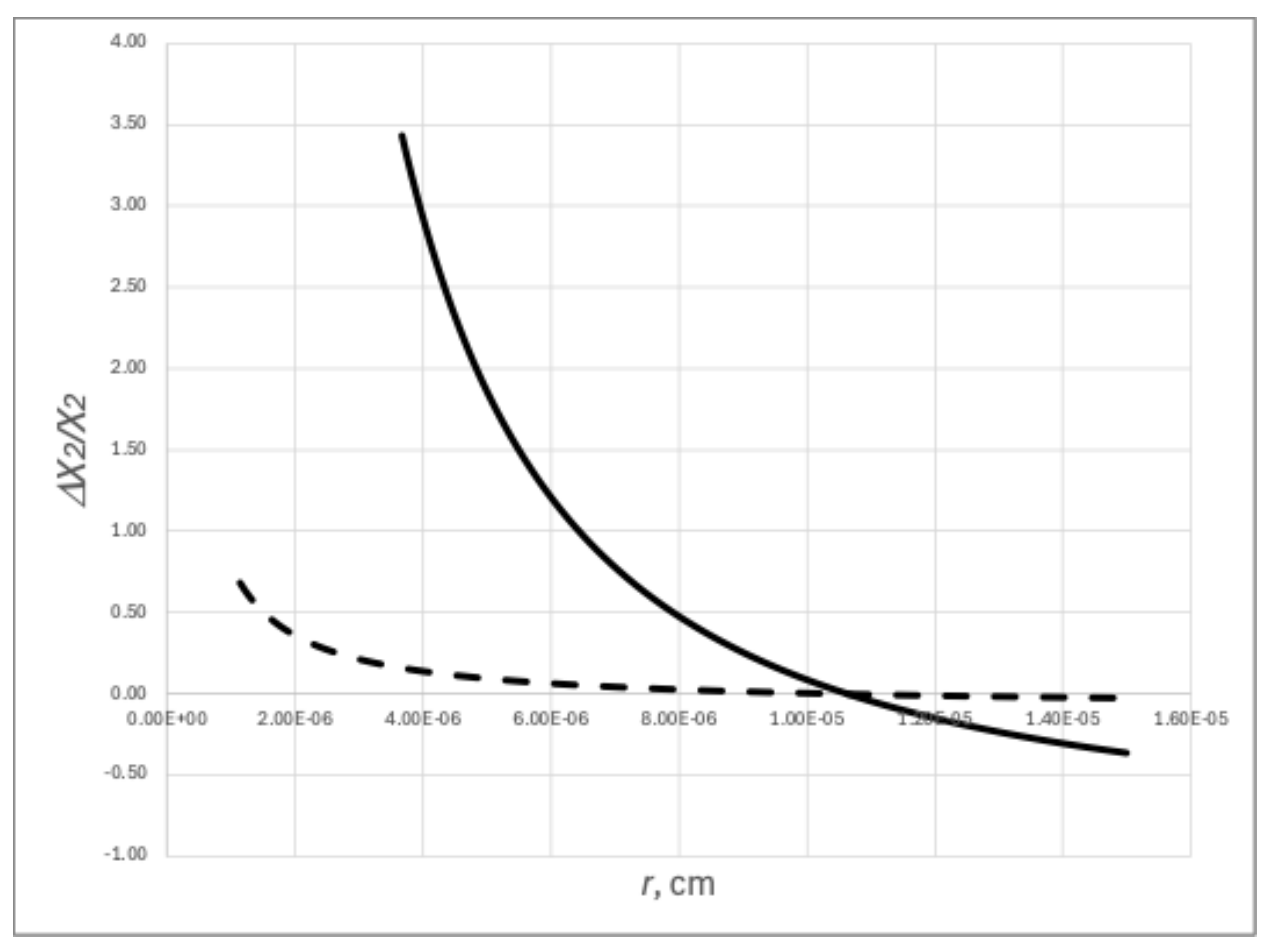

Figure 7. Relative changes in the initial value of the initial mole fraction $x_2$ during the transition to lock-in of LSW plotted versus the final particle size. The same set of parameters as in Figure 6; Dashed line: $x_{02}$ = 0.1, $L_1$ = 11.
; continuous line: $x_{02}$ = 0.1, $L_1$ = 1.

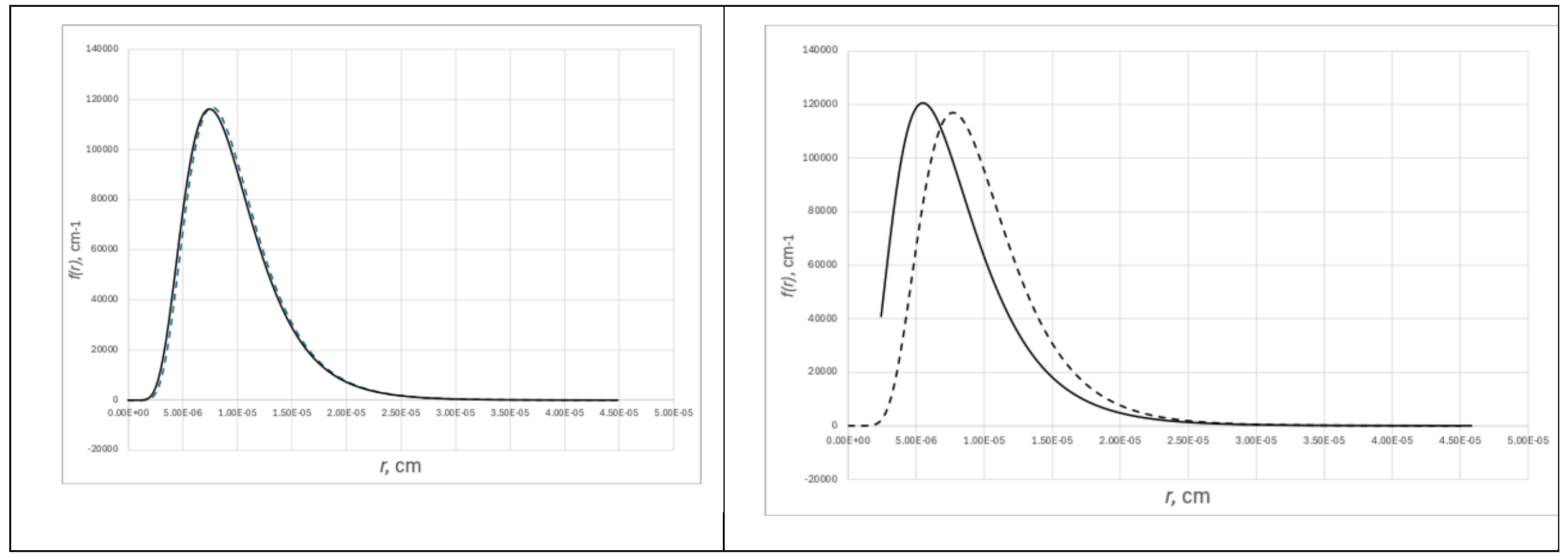

Figure 8. Evolution of the initial log-normal distribution ($\mu$= -0.0957, $\sigma$= 0.408), Ref [10] during the transition to lock-in. Left: $x_{02}$ = 0.1, $L_1$ = 11, $\frac{r_e}{\bar{r}} = 3.50$.; right: $x_{02}$ = 0.01,$L_1$ = 1, $\frac{r_e}{\bar{r}} = 3.50$. The number average radius $\bar{r}$= 100 nm.

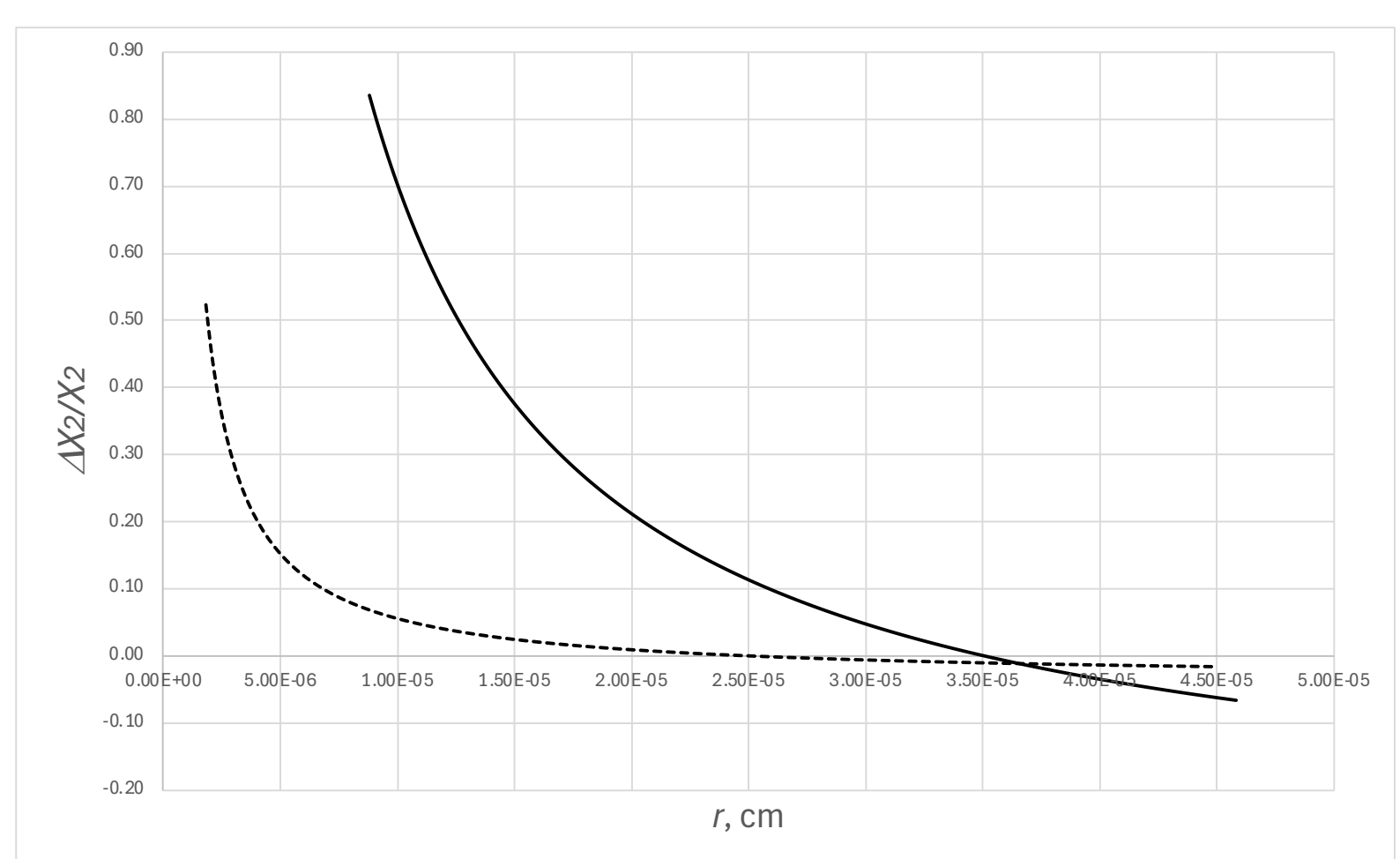

Figure 9. Relative changes in the initial value of the initial mole fraction $x_2$ during the transition to lock-in of lognormal distribution plotted versus the final particle size for lognormal . The same set of parameters as in Figure 8; Dashed line: $x_{02}$ = 0.1, $L_1$ = 11; continuous line: $x_{02}$ = 0.01, $L_1$ = 1.

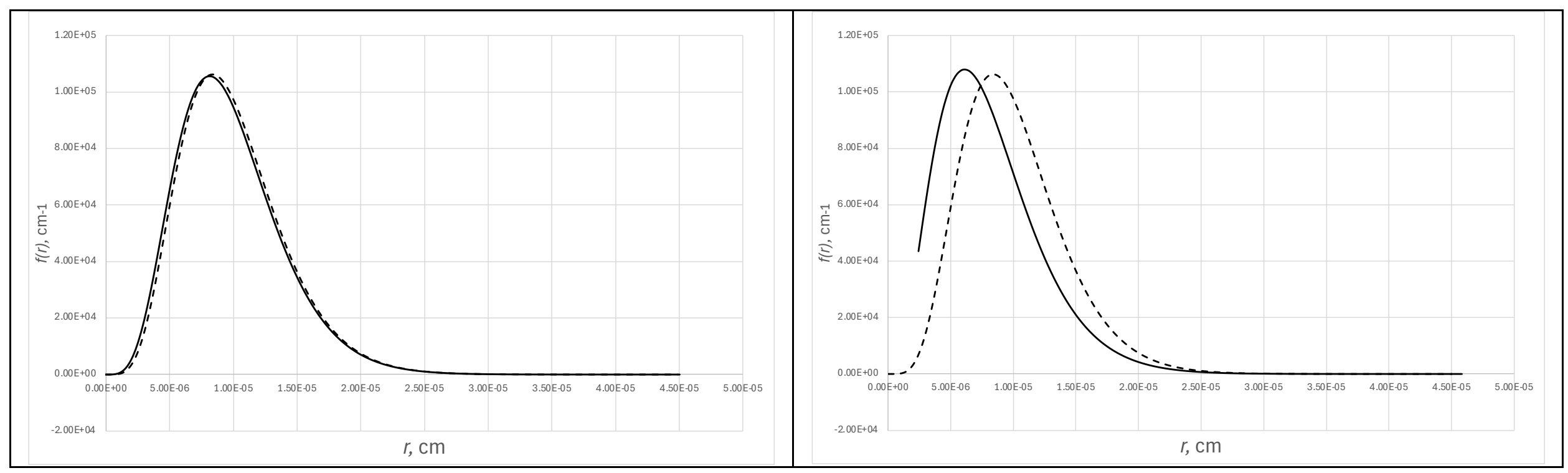

Figure 10. Evolution of the initial Gamma distribution ( n=6.138 ), Ref [10] during the transition to lock-in. Left: $x_{02}$ = 0.1,  $L_1$ = 11, , $\frac{r_e}{\bar{r}} = 3.50$ ; right:  $x_{02}$ = 0.01,  $L_1$ = 1,  $\frac{r_e}{\bar{r}} = 3.50$ and the number average radius of  $\bar{r}$= 100 nm.

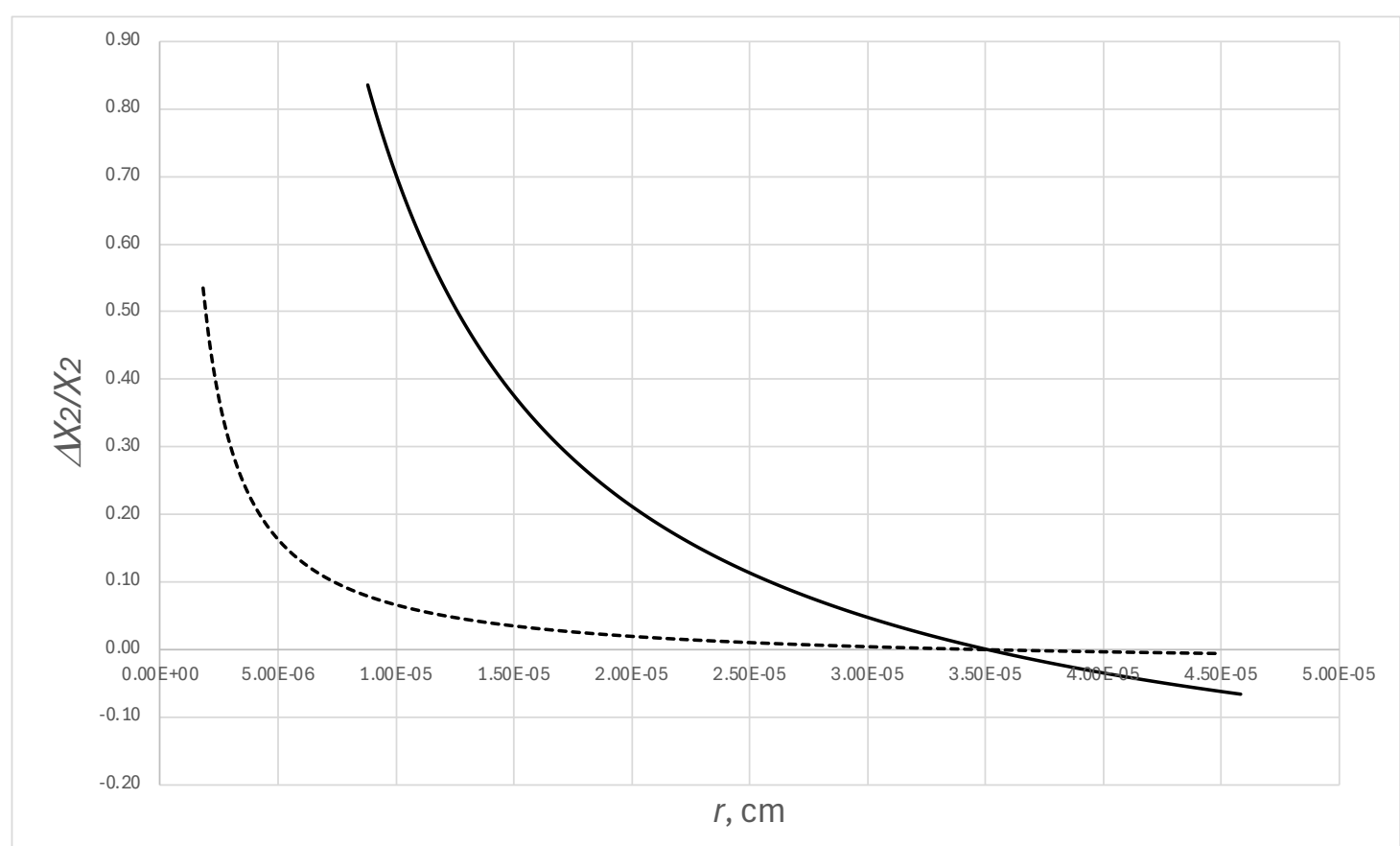

Figure 11. Relative changes in the initial value of the initial mole fraction $x_2$ during the transition to lock-in of Gamma distribution  plotted versus the final particle size for gamma distribution. The same set of parameters as in Figure 10; Dashed line: $x_{02}$ = 0.1, $L_1$ = 11; continuous line: $x_{02}$ = 0.1, $L_1$ = 1.

Table 1. Fraction of the particle size distribution with $\left|\frac{\Delta x_2}{x_{02}}\right| < 0.1$ in the lock-in state

| Distribution | $L_1$ | Volume fraction of the distribution with $\left|\frac{\Delta x_2}{x_{02}}\right| < 0.1$ in the lock-in state |
|---|---|---|
| Uniform | 22 | 100% |
| LSW | 11 | 99.8% |
| Log-normal ($\mu$= -0.0957, $\sigma$= 0.408), Ref [10] | 11 | 96.3% |
| Gamma (n=6.138), Ref [10] | 11 | 97.7% |

**Insoluble Second Component, $L_1$<<1, Bimodal Distribution Development**

Figures 12,13 demonstrate the development of bimodal distributions in case of the low concentration of the second component, $L_1$ = 0.1. The system was assumed to have a log-normal distribution initially; it is predicted to split into the fraction of fine particles, peaked at about 20 nm, and a particle size fraction that coarsens with time. The latter fraction is predicted to follow the classical cubic kinetics with the particle size distribution tending to the classical LSW attractor. The overall evolution of the particle size distribution during this ripening process is peculiar because the total number of particles remains constant in the process; accordingly, when plotted as a number-based distribution, the larger particle size fraction becomes barely visible with time; however, it contains most of the original material, volume-wise. The details and assumptions involved in the predicted particle size distribution evolution are described in the Appendix.

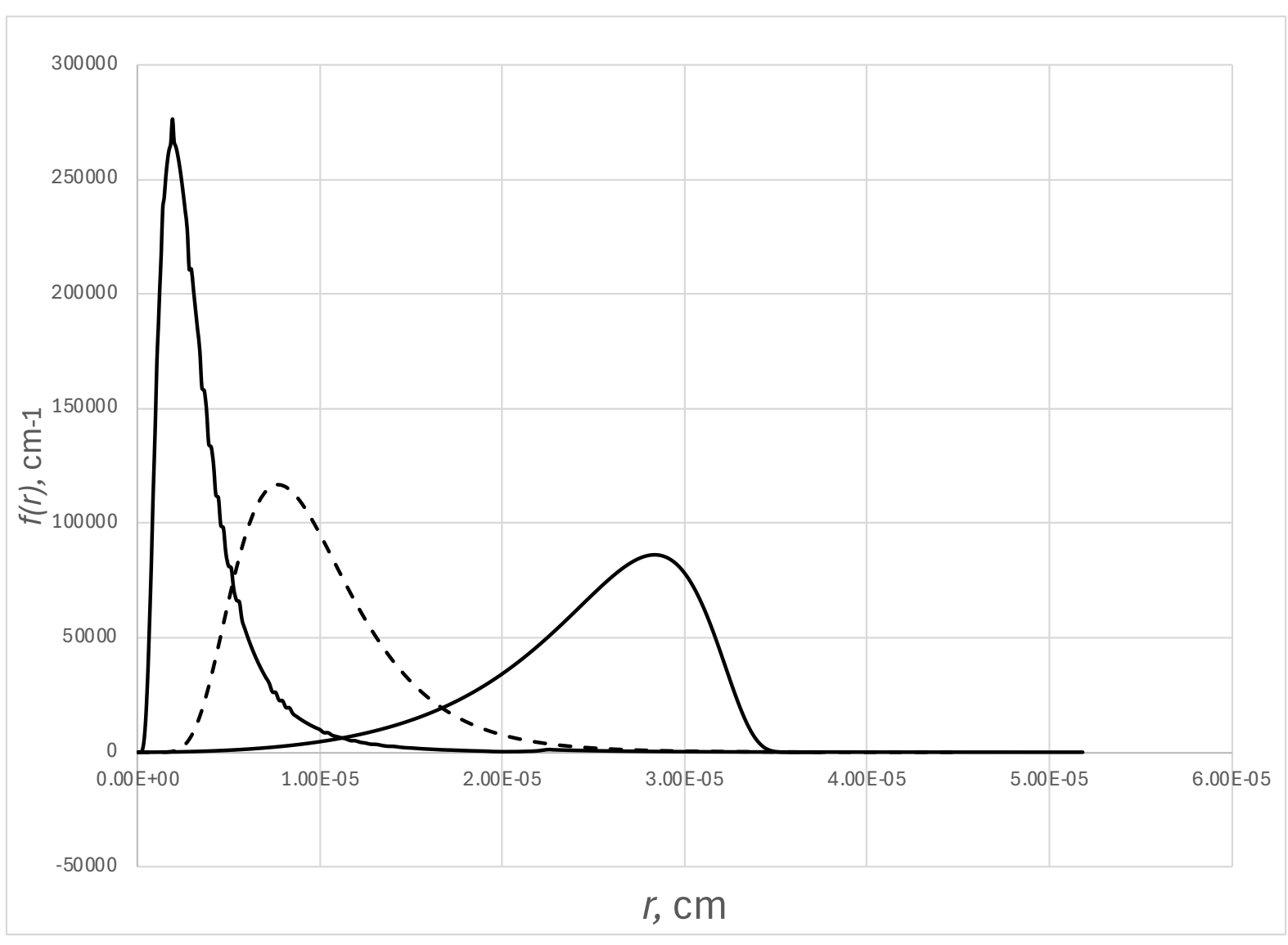

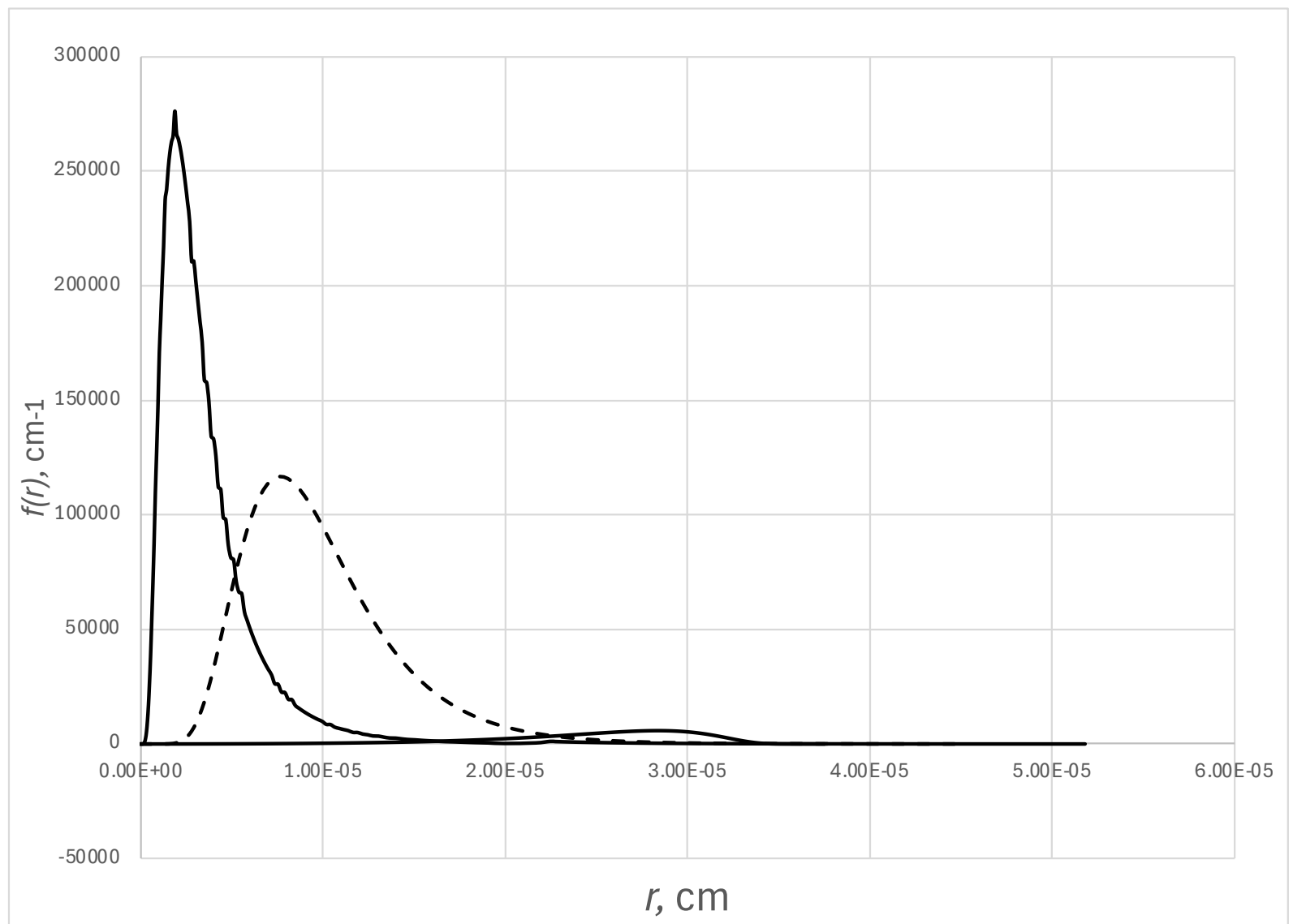

Figure 12. Evolution of a lognormal distribution $\mu$= -0.0957, $\sigma$= 0.408, Ref. [10], with the initial average radius of 100 nm (dashed line) to a bimodal distribution, at the time moment when the average size of the ripening fraction reaches 250 nm. The other parameters: $x_{02}$ = $10^{-3}$, $\alpha_1$ = $10^{-7}$ cm, $L_1$ = 0.1. The 'fines' fraction is predicted to have a maximum at 20 nm. Top: the final state plots are not normalized for visual purposes; bottom: the same plots normalized to the total integral of 1.

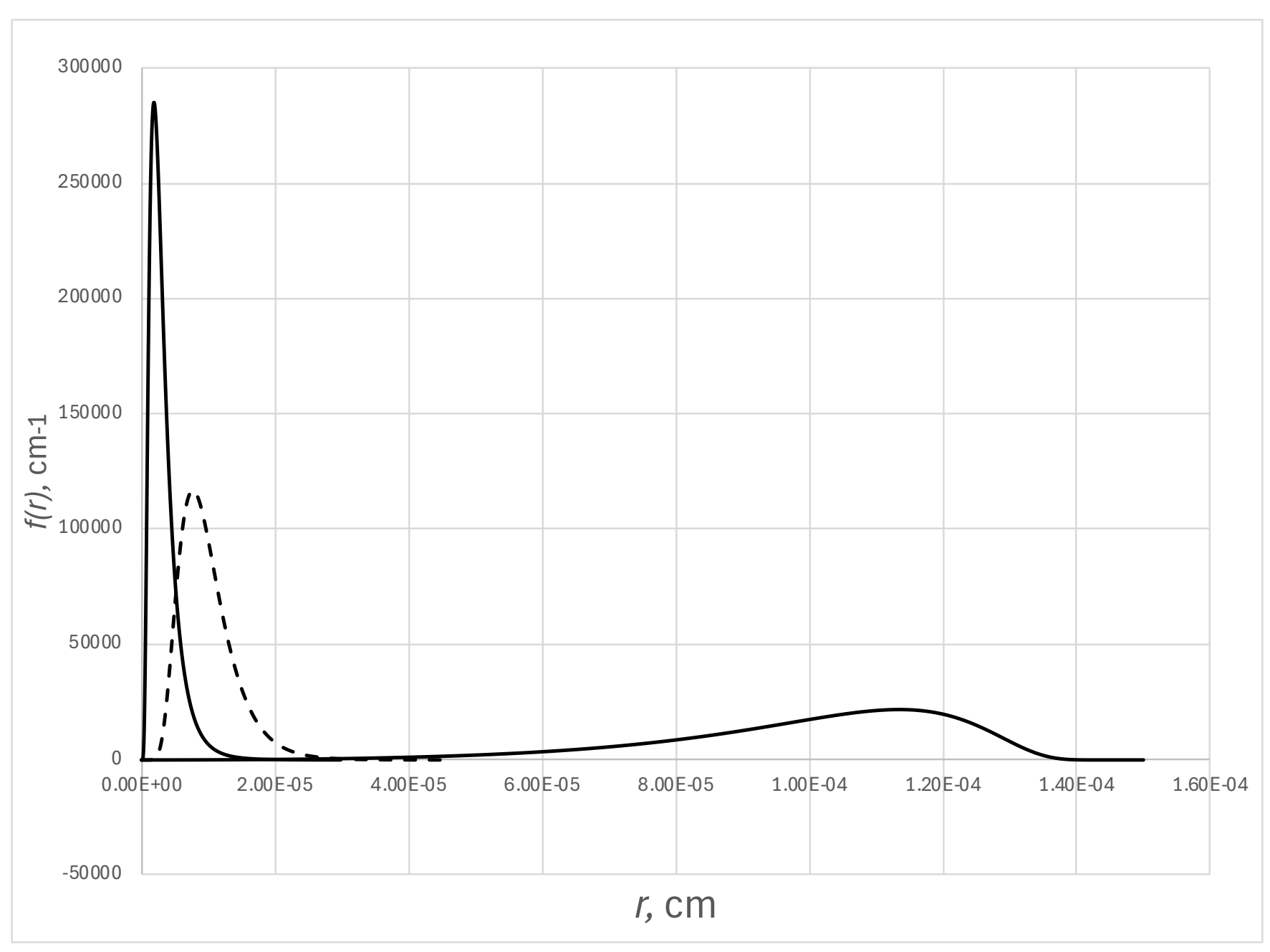

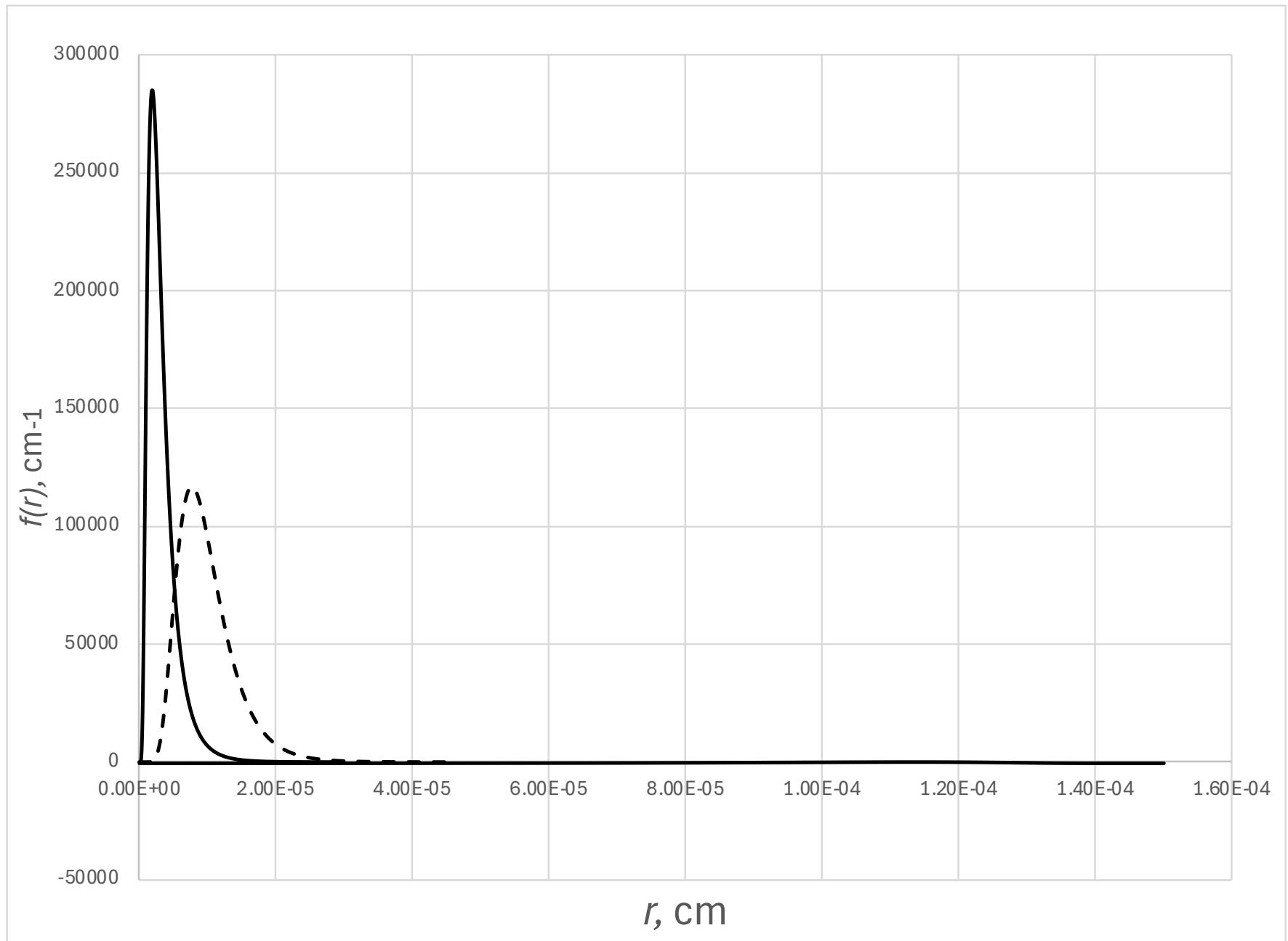

Figure 13 Later stage of evolution of a lognormal distribution, $\mu$= -0.0957, $\sigma$= 0.408 , Ref. [10], with the initial average radius of 100 nm (dashed line) to a bimodal state, at the time moment when the average size of the ripening fraction reaches 1000 nm. The other parameters: $x_{02}$ = $10^{-3}$, $\alpha_1$ = $10^{-7}$ cm, $L_1$ = 0.1. The 'fines' fraction shows a maximum at 19 nm. Top: the final state plots are not normalized for visual purposes; bottom: the same plots

normalized to the total integral of 1; the ripening fraction becomes invisible due to the overlap with X axis.

**Lock-In State. 2. Case of Sparingly Soluble Second Component**

Up to this moment, the second component was considered to be completely insoluble in the medium. We now relax this condition and allow the second component to diffuse, with the analysis closely following our previous papers [2], [4].
For this, we continue with the analysis of the chemical potential changes. We choose the macrophase of the original composition ($x_2 = x_{02}$; $x_1 = x_{01}$) as the reference point for the chemical potentials; the solubilities in this state are equal to $C_{\infty 1}$ and $C_{\infty 2}$ . The excess chemical potentials $\Delta\mu_{1,2}$ of the Components 1 and 2 between the drop and the macrophase are shown below:

$$\Delta\mu_1 = \mu_1(x_1, \Delta p) - \mu_1(x_{01}, \Delta p = 0) \quad (23)$$
$$\Delta\mu_2 = \mu_2(x_2, \Delta p) - \mu_2(x_{02}\ \Delta p = 0) \quad (24)$$

At time zero, just after the emulsion was formed, all the difference comes only from the excess Laplace pressure; however, with time, there is also a contribution from the changes in $x_1$ and $x_2$, which we call the Raoult effect contribution. The equilibrium concentrations at the surface of the drop can be evaluated from these values of the excess chemical potentials:

$$RTln\left(\frac{C_1}{C_{\infty 1}}\right) = \Delta\mu_1 \quad (25)$$

$$RTln\left(\frac{C_2}{C_{\infty 2}}\right) = \Delta\mu_2 \quad (26)$$

and the concentration differences of the mass transfer equations are equal to:

$$\Delta C_1 = C_{\infty 1}\frac{\Delta\mu_1}{RT} - C_{M1}(t) \quad (27)$$

$$\Delta C_2 = C_{\infty 2}\frac{\Delta\mu_2}{RT} - C_{M2}(t) \quad (28)$$

Note again that $C_{\infty 1}$ and $C_{\infty 2}$ are different from $C^*_{\infty 1}$and $C^*_{\infty 2}$ of the individual components because the components reduce the solubility of each other in the medium due to Raoult effect.
We consider both chemical potentials as a function of $x_2$ as the independent variable, and the excess pressure $\Delta p$ of Laplace origin. As we are interested in very small changes in composition *and* pressure, we can expand the chemical potentials in series:

$$\Delta\mu_1 = \frac{\partial\mu_1}{\partial x_2}\Delta x_2 + \frac{\partial\mu_1}{\partial p}\Delta p \quad (29)$$

$$\Delta\mu_2 = \frac{\partial\mu_2}{\partial x_2}\Delta x_2 + \frac{\partial\mu_2}{\partial p}\Delta p \quad (30)$$

Here $\Delta x_2 = x_2 - x_{02}$.

The expansion-in-series is based on the assumption of the smallness of the Laplace effect, and the compensating it Raoult effect. This generally holds when the lock-in parameter is large, $L_1>>1$. Although it breaks down at the very late stage of the particle lifetime, or, which is equivalent, for the small particle size fraction of the continuous distribution, the effect on the overall kintetic is believed to be small, insofar as $L_1>>1$.Considering that $\frac{\partial\mu_1}{\partial p} = V_{m1}$; $\frac{\partial\mu_2}{\partial p} = V_{m2}$ and $\Delta p = 2\sigma/r$, we conclude:

$$\Delta\mu_1 = \frac{\partial\mu_1}{\partial x_2}\Delta x_2 + \frac{\alpha_1}{r}RT \quad (31)$$

$$\Delta\mu_2 = \frac{\partial\mu_2}{\partial x_2}\Delta x_2 + \frac{\alpha_2}{r}RT \quad (32)$$

where

$$\frac{2\sigma V_{m1}}{rRT} \equiv \frac{\alpha_1}{r} \quad (33)$$

$$\frac{2\sigma V_{m2}}{rRT} \equiv \frac{\alpha_2}{r} \quad (34)$$

We continue to call the first term of equations (31, 32 ) the Raoult term, and the second the Laplace term. Note that in both cases,

$$\Delta\mu_1 = \mu_1 - \mu_{\infty 1}(x_{02}) \quad (35)$$

$$\Delta\mu_2 = \mu_2 - \mu_{\infty 2}(x_{02}) \quad (36)$$

that is, the excess chemical potentials are measured with the respect of the state before the dispersion, with the infinite radius of curvature[5]. We now utilize the fact that the changes of the chemical potentials due to composition variations are connected through the Gibbs-Duhem equation:

---

[5] note that the selection of the reference point is now different from what it was in eqns 18 and 19, where the chemical potential of the first component was measured with respect to it as a pure individual component.

$$\frac{\partial\mu_1}{\partial x_2}x_{01} + \frac{\partial\mu_2}{\partial x_2}x_{02} = 0 \qquad (37)$$

After some algebra, we conclude that:

$$\Delta\mu_2 = \frac{2\sigma(x_{01}V_{m1}+x_{02}V_{m2})}{rx_{02}} - \frac{x_{01}}{x_{02}}\Delta\mu_{1e} \qquad (38)$$

We have introduced the average molar volume variable, $\bar{V}_m = x_{01}V_{m1} + x_{02}V_{m2}$ . Equation 38 then can be rewritten as:

$$\Delta\mu_2 = \frac{2\sigma\bar{V}_m}{rx_{02}} - \frac{x_{01}}{x_{02}}\Delta\mu_{1e} \qquad (39)$$

As Ostwald ripening in the system will proceed, $\Delta\mu_{1e}$ value becomes a function of time and is expected to drift down as the average size increases and the supersaturation decreases; note however that it does not depend on the radius of an individual particle explicitly and equally applies to all the particles in the system. On the other hand, the first term of eqn (39) is an explicit function of the particle size *r* and determines supersaturation of each individual particle. We now convert eqn (39) to the dependence of the solubility of the second component in the medium on the particle size, in lock-in state, and expand the exponents in series:

$$C_2 = C_{\infty 2}exp(\frac{\Delta\mu_2}{RT}) = C_{\infty 2}\left(1 + \frac{2\sigma\bar{V}_m}{RTrx_{02}}\right)\left(1 - \frac{x_{01}\Delta\mu_{1e}}{x_{02}RT}\right) \qquad (40)$$

We consider the third multiplier as a small radius-independent correction to the solubility that decreases with time, and ignore it going forward; the equation then simplifies to:

$$C_2 = C_{\infty 2}\left(1 + \frac{2\sigma\bar{V}_m}{RTrx_{02}}\right) \qquad (41)$$

The equation has a form of a regular Kelvin equation for the solubility of a one-component system, with the molar volume replaced by the combination $\frac{\bar{V}_m}{x_{02}}$. Thus, the driving force of the dissolution, that is, the difference in solubilities between the particle with the radius *r* and the critical radius, $r_c$ (which can be introduced the same way as for the one-component system) is equal to:

$$\Delta C_2 = C_{\infty 2}\frac{2\sigma\bar{V}_m}{RTx_{02}}\left(\frac{1}{r} - \frac{1}{r_c}\right) \qquad (42)$$

The next step is to relate the solubility of the second component form the pure state $C^*_2$ to the solubility from the mutual solution with the first component, $C_{\infty 2}$. Utilizing the classical solution theory, we note that:

$$\Delta\mu_{2s} = RTln\frac{C^*_{\infty 2}}{C_{\infty 2}} = RTln\frac{1}{x_{02}\gamma_2} \quad (43)$$

or

$$\frac{C^*_{\infty 2}}{C_{\infty 2}} = \frac{1}{x_{02}\gamma_2} \quad (44)$$

Here on the left is the difference of the chemical potentials in the solution in the medium, and on the right, in the solution of the dispersed phase. We have 1 in the numerator on the right-hand side because for the pure Component 2, $x^*_2 = 1$ and the activity coefficient $\gamma_2$=1. The final version of the growth law equations is:

$$\Delta C_2 = \frac{2\sigma\bar{V}_m C^*_{\infty 2}\gamma_2}{RT}\left(\frac{1}{r} - \frac{1}{r_c}\right) \quad (45)$$

$$J_2 = 4\pi r D_2 C^*_{\infty 2}\gamma_2\bar{\alpha}\left(\frac{1}{r_c} - \frac{1}{r}\right) \quad (46)$$

where $\bar{\alpha} = \frac{2\sigma\bar{V}_m}{RT}$.

We remind ourselves now that the ripening is delimited by the diffusion of the second, less soluble component; the first component just adjusts its concentration constantly to stay in the lock-in state with respect to the second. In this process, we assume that the composition of the drops stays constant 'most of the time' and the relative changes in the volume fractions of the components are negligible; that is, during the lifetime of a particle, 'most of the time', $\frac{\Delta x_2}{x_{02}} << 1$ and $\frac{\Delta\phi_2}{\phi_{02}} << 1$ . As it follows from the discussion of the previous sections, $L_1$>>1 is required for this to hold, with $L_1$ >10 being a good starting point for some of the initial distributions.

To keep the volume fractions constant, the total volume constantly adjusts itself to keep the volume fraction constant, that is, the following condition needs to be met:

$$\frac{dV}{dt} = \frac{d}{dt}\left(\frac{4}{3}\pi r^3\right) = \frac{J_2}{\phi_{02}} \quad (47)$$

and we determine the velocity of the particles in the space of sizes as:

$$\frac{dr}{dt} = \frac{D_2\gamma_2 C^*_2\bar{\alpha}}{\phi_{02}r^2}\left(\frac{r}{r_c} - 1\right) \quad (48)$$

which is nearly the same equation as in the single component case of LSW theory, eqn (12). As the only difference between these two rate equations is the constant, size independent factor, we conclude that the whole machinery of the Ostwald ripening theory for the single component case can be leveraged; the same distribution function attractor is predicted, and the ripening rate will change to:

$$w = \frac{d(\bar{r}^3)}{dt} = \frac{8C_2^* \gamma_2 \sigma \bar{V}_m D_2}{9\phi_{02} RT} \tag{49}$$

Here the $\frac{d(\bar{r}^3)}{dt}$ is the rate of increase in the number-averaged radius cubed, per the original LSW theory notation [7, 8]. Just as in the case of the classical LSW theory, Equation (49) is predicted to hold at the late asymptotic stage of the process, after the LSW attractor distribution is established.

Per this equation, the rate is predicted to increase as the volume fraction of the less soluble component decreases, in a hyperbolic fashion. Eventually, it will reach the value for the first, more soluble in the medium component, and further decrease in $\phi_{02}$ should have no effect on the rate. Thus, for small values of $\phi_{02}$, equation (49) breaks down and the rate must tend to:

$$w = \frac{d(\bar{r}^3)}{dt} = \frac{8C_1^* \sigma V_{m1} D_1}{9RT} \tag{50}$$

One of the challenges is to connect these two kinetic regimes into a single equation. If such an equation would be found, it will cover the whole range of compositions, from $\phi_{02}$ = 0 to 1. We also aspire to include the possible deviations of the solution from ideality into this equation. Following the arguments of our previous paper of this series [4] , one may use the extrapolatory equation:

$$\frac{\bar{V}_m}{w} = \frac{\phi_{01} V_{m1}}{w_1 \gamma_1} + \frac{\phi_{02} V_{m2}}{w_2 \gamma_2} \tag{51}$$

or, equivalently,

$$w = \frac{8\sigma \bar{V}_m}{9RT} \left( \frac{\phi_{01}}{C_{01}^* D_1 \gamma_1} + \frac{\phi_{02}}{C_{02}^* D_2 \gamma_2} \right)^{-1} \tag{52}$$

If $\phi_{02}$ is close to1, the second term in paratheses is dominating, and Eqn (49) is recovered; conversely, if $\phi_{02} \ll 1$ and $\phi_{01}$ is close to 1 (which also means that $\gamma_1$ is close to 1), Eqn (50) is recovered.

If  the solution of the components in each other is ideal, the equation (52)  reduces to:

$$w = \frac{8\overline{\sigma V}_m}{9RT}\left(\frac{\phi_{01}}{C^*_{01}D_1} + \frac{\phi_{02}}{C^*_{02}D_2}\right)^{-1} \qquad (53)$$

Finally, if the molar volumes of the components are nearly identical, we get:

$$w = \frac{8\sigma V_m}{9RT}\left(\frac{\phi_{01}}{C^*_{01}D_1} + \frac{\phi_{02}}{C^*_{02}D_2}\right)^{-1} \qquad (54)$$

which reduces to the extrapolatory equation of our previous paper, Ref. [2]:

$$w = \left(\frac{\phi_{01}}{w_1} + \frac{\phi_{02}}{w_2}\right)^{-1} \qquad (55)$$

The result of this paper is therefore an extension of our previous paper's result, which accounts for the difference in the molar volume of the components and accounts for the fact that the mutual solution of the components can be non-ideal. Also, it can be arrived by a simple scaling argument as based on the result of Ref. [4] for the dissolution of a single particle.

## Summary and comparison with experiment

In this paper, we revisited the problem of Ostwald ripening in systems with two-component dispersed phase, extending the results of the previous studies [2], [3], [4]. In case of a completely insoluble second component, depending on the value of the lock-in parameter $L_1$, the system either enters an equilibrium state, with nearly no changes to the particle size ($L_1$>>1) or decomposes into two fractions, one undergoing the classical ripening, and the other catching all the small particles that cannot disappear completely. This distribution is predicted to remain bimodal over  time ($L_1$<<1). There is also a range  in between, when $L_1$ is close to 1. The behavior there is strongly dependent on the initial distribution function.  For a narrow initial distribution, the system still can enter a lock-in. Thus, at 1/3<$L_1$<1, an initially monodispersed system may come to a metastable lock-in [2, 3], with no further ripening.  For  broad initial distributions, a part of the distribution may undergo the lock in, whereas another fraction will ripen; the scenario will be similar to be described  for  $L_1$<<1 case,  with the development of a bimodal distribution, with the main and 'fine' fractions substantially overlapping  with each other. The summary of the possible regimes is given below in Table 2 and is very much in line with the earlier work on this subject [2, 3], with some minor extensions.

Table 2. Possible Ostwald ripening scenarios for two-component systems with insoluble second component [2, 3]

| Regime | $L_1$ value | Comments |
|---|---|---|
| Full lock-in | $L_1$>>1 | The system comes to equilibrium and the mass transfer stops. The particle size distribution undergoes only a minor change and remains close to the initial distribution; $\Delta r/r$<<1. The relative changes in composition are also small, $\Delta x_2/x_{02}$ <<1, except for the range of the small particles, r<<$\bar{r}$, where $\Delta x_2/x_{02}$ is no longer small. The volume fraction of particle range for which the condition $\Delta x_2/x_{02}$ <<1 is violated gets progressively smaller as $L_1$ increases. |
| Partial lock-in | $L_1$ ~1 | The behavior is strongly dependent on the initial distribution function. For narrow initial distribution, the system still can enter a lock-in, although $\Delta r/r$ and $\Delta x_2/x_{02}$ are no longer small. Thus, at 1/3<$L_1$<1, an initially monodispersed system may come to a metastable lock-in [2, 3], with no ripening. For broad initial distributions, a part of the distribution may undergo the lock in, whereas another fraction will ripen; the scenario will be similar to be described for $L_1$<<1 case, with the development of a bimodal distribution, with the main fraction and the fraction of 'fines' substantially overlapping with each other. |
| No lock in | $L_1$<<1 | The system undergoes ripening as if no additive were present. The small particles however do not disappear but accumulate as the fraction of 'fines'. The particle size distribution becomes bimodal. |

Numerical calculations were conducted to model this behavior for various initial distribution functions of log-normal type, as well for the LSW attractor function, to understand the degree of the change in distribution during the transition of the system into the lock-in.

The condition of zero solubility of the second component is then relaxed to low but final solubility. In case when the lock-in number $L_1$ is high, $L_1$ >>1, the changes in the composition of the ripening particles remain small during most of their lifetime and the condition $\frac{\Delta x_2}{x_{02}} << 1$ holds, except for the very last moment of dissolution. The latter does not contribute substantially to the kinetics, which mostly follows the classical cubic pattern; the arguments here are very similar to the ones made by us before in the theory of the dissolution of the single particle, Ref [4]. The exact threshold value for $L_1$ is difficult to pinpoint as it depends on the degree of polydispersity of the initial particle size distribution; however, $L_1$>10 may be a good starting point for the estimate.
An improved extrapolatory equation for the ripening rate is derived, which accounts for the difference in the molar volumes and diffusion coefficients of the diffusing species; it also explicitly includes the solution nonideality. The whole complex machinery of LSW theory can be fully leveraged in this case; specifically, the particle size distribution is predicted to be nearly identical to LSW attractor, as the changes in distribution as caused by the lock-in are minor. As in LSW theory, the attractor establishes itself at the late stages of ripening, in the asymptotic regime.
Although the kinetic model heavily relies on the approximation $\frac{\Delta x_2}{x_{02}} << 1$, the approximation does not hold for all the particles in the continuous distribution; it always has to break down for the fraction of small particles; in other words, there is always a fraction of particles in the continuous distribution for which it breaks down, no matter how large $x_{02}$ is. To put this statement in an equivalent form, there is always a stage of the dissolution of a particle, in the very end of its lifetime, when this approximation has to break down. The argument that is made in this paper is similar to the one made in the previous paper of this series [4]; although this is true, the overall effect on the kinetic of dissolution is small as the duration of this stage is very small -- the more so, the larger is the $L_1$ number. To a degree, the approximation here is similar to the linearization of Kelvin equation in LSW theory: whereas the equation can be linearized for the average radius of the distribution, the linear approximation breaks down in the very end of the particle lifetime, but this fact does not affect the kinetics much.
One case that was not specifically addressed in this paper is the scenario when $L_1$<1 but the solubility of the second component is not zero. A bimodal distribution is predicted to emerge in this case as well at first, but the fraction of 'fines' is expected to show some dynamics to it; thus, with time it is predicted to gradually transfer back to the main fraction. It can also Ostwald-ripen within itself, but this process is expected to be much slower. Developing this model may be an object of a separate study.
After the theory of ripening in two-component mixtures were proposed [2], it was validated in ripening of hydrocarbon [11] and fluorocarbon [12,13] emulsions with two-component dispersed phase. The accuracy of the measurements and the degree of variations in the molar volumes of the components is not sufficient to confirm that the new version of the equation eqn (51) proposed in this paper, will provide an improvement over the old, eqn (55), to which the data were originally mapped to; therefore, more experiments are needed to resolve this. The development of a lock-in state in with ripening system was also confirmed in a very nice study [13] in which the composition of emulsion drops of a different size was analyzed.
As far as the emergence of the bimodal distribution concerned, it was observed experimentally on a qualitative level in alkane emulsions with very small amounts of stabilizing additives of higher alkanes [11]. Broadly speaking, the existence of 'fines' fraction in emulsion manufacturing has been known but is not specifically associated to the ripening phenomena [14]. The fractions of 'fines' are also known to emerge in emulsion and suspension polymerization, which, in the view of this paper, may be associated with ripening of monomer emulsions in presence of small impurities of partially polymerized oligomers.
It should be noted that bimodal distributions are also quite common in aerosol science, specifically in moisture clouds, as water condenses on the crystals of salts over oceans [15]. Although the phenomenon is not normally associated with ripening, it could be a factor to consider.

## Concluding remarks

We would like to finish the paper by some physical insights, as applied to O/W emulsions. Having an additive with a low solubility in water always helps to stabilize ripening; In this sense, higher alkanes, and fats C16 -C18 effectively stabilize emulsions against ripening in the 100 nm range. It should be noted that increasing the chain length further is not expected to bring a benefit as the solubility in water is already quite low. The opposite may happen, as the molar fraction $x_2$ decreases because of the growth in molecular weight and the lock-in criterion will be more difficult to meet. That is, polymeric hydrophobic additives are not expected to be good stabilizers against ripening.
Another approach that a formulator can take is to reduce the interfacial tension. Not only it will proportionally decrease the ripening rate; it will also enable the lock-in state criterion $L_1 >> 1$ at the smaller amount of the additive. Thus, whereas interfacial tensions for common surfactants at the oil -water interface are in 10 dyn/cm range, the use of the 'balanced' surfactants or surfactant mixtures can reduce the interfacial tension by additional 3-4 orders of magnitude, to ~$10^{-2}$-$10^{-3}$ dyn/cm, thereby reducing $\alpha$ to $10^{-3}$ nm, and enabling the stabilizing effect of very small amounts of additives even for few nanometer particles.

### Acknowledgments

Comments of Prof. Håkan Wennerström on the manuscript are appreciated.

## References

[1] W.I. Higuchi, and J. Misra, "Physical Degradation of Emulsions Via the Molecular Diffusion Route and the Possible Prevention Thereof", *Journal of Pharmaceutical Sciences*, Vol. 51, no. 5, pp. 459-466, 1962.
[2] A.S. Kabalnov, A.V. Pertzov, and E.D. Shchukin," Ostwald ripening in two-component disperse phase systems: Application to emulsion stability", *Colloids and Surfaces*, Vol 24, no. 1, pp. 19-32, 1987.
[3] A. J. Webster and M. E. Cates, "Stabilization of Emulsions by Trapped Species", *Langmuir* vol 14, no 8, pp. 2068-2079, 1998.

[4] A. Kabalnov, "Dissolution of a two-component drop onto macrophase due to surface tension effect", *Open Transport,* 2026, accepted.
[5] I. M. Lifshitz and V. V. Slezov, “On diffusional decay kinetics of supersuturated solid solution,” *Zh. Eksp. Teor. Fiz*. vol 35, pp. 497–505 (1958).
[6] C. Wagner, “Theorie der Alterung von Niderschlagen durch Umlösen (Ostwald Reifung),” *Z. Electrochem*. vol 65, pp. 581–591 (1961).
[7] J. Crank, *The mathematics of diffusion*. Oxford university press, 1979.
[8] A. Kolmogorov. "On the log-normal distribution of the fragment sizes under breakage". *Dokl. Akad. Nauk SSSR 31*, pp 99– 101, 1941.
[9] Villermaux, E. "Fragmentation". *Annu. Rev. Fluid. Mech.* 2007, 39, 419– 446.
[10] M. L’Estimé, M. Schindler, N. Shahidzadeh, and D. Bonn, "Droplet Size Distribution in Emulsions" *Langmuir* vol 40, no 1, pp 275-281, 2024.
[11] A.S. Kabalnov, A.V. Pertsov, Y.D. Aprosin, and E.D. Shchukin , "Influence of nature and composition of disperse phase on stability of O/W emulsions against Ostwald ripening" *Colloid Journal of the USSR*, vol 47, no 6, pp 898-903 (1985).
[12] A.S. Kabalnov, Yu. D. Aprosin, O.B. Pavlova-Verevkina, A.V. Pertsov, and E.D. Shchukin, "Influence of nature and composition of disperse phase in emulsions of perfluoroorganic compounds on the kinetics of their coarsening" ,*Colloid Journal of the USSR* vol. 48, no 1, pp 20-24 (1986).
[13] J. G. Weers, and R. A. Arlauskas. "Sedimentation field-flow fractionation studies of Ostwald ripening in fluorocarbon emulsions containing two disperse phase components." *Langmuir* vol 11, no 2, pp 474-477, 1995.
[14] M. Li, C. Liu, C. Liang, C. Liu, and J. Li, "Study of Bimodal Drop Size Distributions of Emulsion", *Journal of Dispersion Science and Technology*, vol *35, no* 3, pp 397–402, 2014.
[15] Korolev, A. "A study of bimodal droplet size distributions in stratiform clouds". *Atmospheric Research*, vol 32, pp 143-170, 1994.

## Appendix

**Calculations of the particle size distribution changes during lock-in transition for insoluble second component.** Numerical calculations were conducted to evaluate the change in the particle size for different initial particle size distributions. The calculations were conducted under the assumption of an ideal solution of the components in each other, and equal molar volumes of the components forming the mixture. The components were assumed to be soluble in each other over the whole range of compositions from $x_2$ = 0 to $x_2$ = 1 and the volume of mixing effect was neglected. Accordingly, the molar fraction of the second component was evaluated as:

$$x_2 = x_{02}\left(\frac{r_0}{r}\right)^3 \qquad (56)$$

and the excess chemical potential of the first component was calculated as:

$$\frac{\Delta\mu_1}{RT} = ln(1 - x_2) + \frac{\alpha_1}{r} \qquad (57)$$

For cases of $L_1$>=1, the calculation was initiated at the number-average value of $r_{eq} = \bar{r}$ and the excess chemical potential $\frac{\Delta\mu_1}{RT}(r_{eq})$ was calculated at this value of the radius. The distribution function range was divided on 150 points, and for each radius point $r_0$, the radius of the particle $r$ satisfying the equation (57) was calculated numerically by solving the equation

$$\frac{\Delta\mu_1}{RT} = ln(1 - x_2) + \frac{\alpha_1}{r} = \frac{\Delta\mu_1}{RT}(r_{eq}) \qquad (58)$$

for each point. Particles larger than $r_{eq}$ needed to grow and particles smaller than $r_{eq}$ needed to diminish to meet this condition. On completion, the material balance of the new distribution was conducted. The value of $r_{eq}$ was then increased and the whole procedure repeated till the material balance condition was met. The calculation was therefore iterative with respect to the value of $r_{eq}$. As expected, the equilibrium value of $r_{eq}$ was larger than the number averaged value. Note that the material balance only included the volume of the particles themselves; the amount of the material in the form of solution in the medium was ignored as negligibly small.

For cases of $L_1$<1, the calculation was conducted only for the fraction of fines; the distribution of the ripening fraction was assumed to be the same as the LSW function, which was an oversimplification, as it takes time for the true LSW attractor to develop. The calculation process was similar to the one above, with the difference that the value of the equilibrium excess chemical potential was set as:

$$\frac{\Delta\mu_{eq}}{RT} = \frac{\alpha_1}{\bar{r}_{LSW}} \qquad (59)$$

where $\bar{r}_{LSW}$ is the assumed number average radius of the larger-particle-size ripening fraction, which is also the critical radius that determines the value of the supersaturation, per LSW theory; due to the strong dilution, the presence of the insoluble additive in the ripening larger particle size fraction was neglected. The equilibrium size of the particles of the fines fraction was constrained to be on the left side from the chemical potential maximum (Fig 2). No material balance condition was used in this case as all the excess material was assumed to be absorbed by the ripening fraction.